\documentclass[aps, pra, showpacs, amsmath, preprint, groupedaddress]{revtex4}
\usepackage{graphics}
\usepackage{graphicx}
\usepackage{subfigure}
\usepackage{float}
\usepackage{longtable}
\usepackage{amssymb}
\usepackage{bm}

\begin{document}
\title{Ionization of oriented targets by intense circularly polarized laser pulses: Imprints of orbital angular nodes in the 2D momentum distribution}
\author{C. P. J. Martiny}
\author{M. Abu-samha}
\author{L. B. Madsen}
\affiliation{Lundbeck Foundation Theoretical Center for Quantum
System Research,
 Department of Physics and Astronomy, Aarhus University, 8000 {\AA}rhus C, Denmark.
}

\date{\today}

\begin{abstract}
We solve the three-dimensional time-dependent Schr\"{o}dinger equation for a few-cycle circularly polarized femtosecond laser pulse interacting with an oriented target exemplified by an Argon atom, initially in a $3\text{p}_{x}$ or $3\text{p}_{y}$ state. The photoelectron momentum distributions show distinct signatures of the orbital structure of the initial state as well as the carrier-envelope phase of the applied pulse. Our \textit{ab initio} results are compared with results obtained using the length-gauge strong-field approximation, which allows for a clear interpretation of the results in terms of classical physics. Furthermore, we show that ionization by a circularly polarized pulse completely maps out the angular nodal structure of the initial state, thus providing a potential tool for studying orbital symmetry in individual systems or during chemical reactions.
\end{abstract}


\maketitle

\section{Introduction}
The recent decades have shown substantial progress in strong-field physics with pulsed lasers. Pulses with field strengths equivalent to the field of the Coulomb interaction in ground state atoms and durations of only a few femtoseconds are now available for a large range of wavelengths \cite{Scrinzi, Krausz}. This development has facilitated the opening of exciting new research areas such as attoscience \cite{Scrinzi, Krausz} which, in turn has set entirely new standards for the interrogation of atomic and molecular dynamics.
Strong-field ionization is a process of special relevance and importance within strong-field physics, since ionization triggers other strong-field phenomena, such as high-harmonic generation, the key process for generating coherent attosecond pulses in the XUV regime. In strong-field studies there is also an interest in creating unique quantum targets by molecular alignment and orientation techniques \cite{Stapelfeldt}. Spatial control of molecules is of importance within attoscience, femtochemistry and molecular reactivity among others. Recent experiments not only succeeded in alignment of molecules but also in 3D orientation, thus giving complete control over the spatial orientation \cite{Baumfalk, Sakai, Tanji, Buck, Holmegaard, Nevo}.

The development described above motivates a detailed investigation of strong-field ionization of oriented targets by strong laser pulses with special emphasis on orbital structure, e.g., effects of orbital symmetry. The bulk of experimental and theoretical works within strong-field physics have dealt with linearly polarized laser pulses, and there is now a growing interest in the study of ionization of atoms and molecules by elliptically polarized pulses \cite{Eckle1, Eckle2, Staudte, Magrakvelidze, Mig5}. The effects of angular orbital structure in the ionization by linearly polarized fields have previously been studied for direct electrons \cite{Bisgaard, Bisgaard2} as well as for rescattered electrons \cite{Busuladzic, Meckel}. When it comes to the probing of angular orbital structure through photoelectron angular distributions or momentum distributions, the use of a strong circularly polarized probe rather than a linearly polarized one has, however, two major advantages: (i) In the circularly polarized field, an electron born in the continuum is constantly driven away from the nucleus, due to the polarization. This dynamics minimizes rescattering effects as well as interference between wave packets launched at different instants of time during the driving pulse. Thus, the use of circularly polarized laser pulses entails a cleaner ionization signal with respect to orbital structure. (ii) The polarization plane is two-dimensional, which permits for a more transparent interrogation of angular orbital structure as we shall see in the present work. In fact, it has recently been shown experimentally that strong-field ionization of 3D oriented $\text{C}_{7}\text{H}_{5}\text{N}$ molecules by a circularly polarized field polarized in the nodal planes of the outermost orbitals provides a unique probe of the angular nodal structure \cite{Holmegaard2}.

In this paper, we focus on the case where the laser polarization plane is perpendicular to the nodal plane. We calculate, by solving the time-dependent Schr\"{o}dinger equation (TDSE), the photoelectron momentum distribution for ionization of an Argon atom initially prepared in a $3\text{p}_{x}$ or $3\text{p}_{y}$ state modeling an oriented target with a single nodal plane. The $3\text{p}_{x}$ and $3\text{p}_{y}$ states, in particular, serve as models for investigating ionization of a molecular orbital with $\pi$ symmetry. In the strong near infrared laser field, the ionization is tunneling-like and the photoelectron is born in the continuum at a relatively large distance from the centre-of-mass. After the ionization, the circularly polarized field drives the electron away from the core, which in turn minimizes the importance of the detailed structure of the molecular potential at small distances. The calculations are compared with results obtained using the length gauge strong-field approximation (LG-SFA) \cite{Keldysh, Lewenstein, Kjeldsen2}, which facilitates interpretation using semiclassical theory \cite{Corkum}. Our results show distinct effects of the angular nodal structure of the initial orbital. For example the angular nodal structure of $\text{p}_{x}$, $\text{p}_{y}$ and $\text{d}_{xy}$ orbitals is readily mapped out, showing that strong-field ionization by a circularly polarized laser pulse directly probes the nodal structure.

This paper is organized as follows. In Sec.~\ref{Theory} we briefly review the basic theory behind the computations. In Sec.~\ref{Results} we discuss the results and in Sec.~\ref{Conclusion} we conclude. Atomic units $[\hbar=a_{0}=m_{e}=1]$ are used throughout, unless stated otherwise.

\section{Theory}\label{Theory}
We investigate an oriented target, modeled by an Ar atom initially prepared in a $3\text{p}_{x}$ or $3\text{p}_{y}$ state, interacting with a circularly polarized few-cycle pulse also in the $xy$ plane. The electric
field is defined as $\vec{E}=-\partial_{t}\vec{A}$ with the vector potential $\vec{A}(t)$,
\begin{eqnarray}
\vec{A}(t)=\frac{A_{0}}{\sqrt{2}}f(t)\left(\begin{array}{c}
                      \cos\left(\omega_{0} t+\phi\right) \\
                      \sin\left(\omega_{0} t+\phi\right) \\
                      0
                    \end{array}
\right).
\end{eqnarray}
Here $A_{0}$ denotes the amplitude, $\omega_{0}$ the carrier frequency, $\phi$ the carrier-envelope phase (CEP) and $f(t)=\sin^{2}\left(\frac{\omega_{0}t}{2N}\right)$ the envelope, with $N$ the number of optical cycles. The laser field couples primarily to the least bound electron, which in turn motivates the use of the single-active electron approximation (SAE). Hence, our analysis is restricted to the outermost electron, the remaining being described by an effective potential \cite{Muller}. The photoelectron momentum distribution is given by
\begin{equation}
\frac{\partial^{3} P}{\partial q_{x} \partial q_{y} \partial q_{z}}=|\langle\Psi^{-}_{\vec{q}}(\vec{r})|\Psi(\vec{r},T)\rangle|^{2},
\end{equation}
where $\Psi^{-}_{\vec{q}}(\vec{r})$ is a continuum scattering wave function of asymptotic momentum $\vec{q}$ and $\Psi(\vec{r},T)$ is the wave packet at the end of the pulse, $T$. We calculate the wave packet by expanding in a spherical harmonic basis for the angular part and reduced radial wave functions $f_{lm}(r)$ defined on a radial grid, $\Psi(\vec{r},t)=\sum_{l=0}^{l_{\text{max}}}\sum_{m=-l}^{l}(f_{lm}(r,t)/r)Y_{lm}(\theta,\phi)$, and then solving the TDSE in the velocity gauge using a split-step method. The details are described elsewhere \cite{Muller2, Kjeldsen}. The scattering states entering in Eq. (2) are obtained by solving the time-independent Schr\"{o}dinger equation for the effective potential.

The LG-SFA \cite{Keldysh, Lewenstein}, which completely neglects the Coulomb interaction in the final state as well as all intermediate states, offers a simple formula for the momentum distribution
\begin{equation}
\frac{\partial^{3} P}{\partial q_{x} \partial q_{y} \partial q_{z}}=\left|\int^{T}_{0}\langle\Psi^{V}_{\vec{q}}(\vec{r},t)|\vec{E}\cdot\vec{r}|\Psi_{i}(\vec{r},t)\rangle dt \right|^{2},
\end{equation}
where $\Psi^{V}_{\vec{q}}$ is a Volkov wave function with asymptotic momentum $\vec{q}$ and $\Psi_{i}$ is the initial state. In the evaluation of Eq. (3) it is accurate to use the asymptotic form, $\Psi_{i}(\vec{r},t)=\sum_{lm}C_{lm}r^{\nu-1}\exp(-\kappa r)Y_{lm}(\theta,\phi)\exp(iI_{p}t)$, of the initial state, since ionization primarily occurs at large distances from the nucleus. Here $I_{p}$ denotes the ionization potential, $\kappa=\sqrt{2I_{p}}$ and $\nu=Z/\kappa$ with $Z$ the charge of the residual ion. The spatial integration in Eq. (3) can be performed analytically, within this approximation, while the 1D time integral is evaluated effectively using the saddle-point approximation \cite{Milosevic}, which is accurate in the parameter regime considered in this paper \cite{Mig2, Mig4}.
\section{Results}\label{Results}
\subsection{Imprints of nodal planes in momentum distributions}
Figure 1 shows the TDSE results for the momentum distributions, in the $xy$ plane of polarization, for the (a, b) $3\text{p}_{x}$ and (c, d) $3\text{p}_{y}$  initial states of Ar, probed by an 800 nm 3-cycle laser pulse with peak intensity $I=1.06 \times 10^{14}\text{W}/\text{cm}^{2}$ and CEP (a, c) $\varphi=-\pi/2$ , (b, d) $\varphi=0$. The distributions are calculated using a 4096 points radial grid extending to $r_{\text{max}}=400$, maximum angular momentum $l_{\text{max}}=40$, and a time-step of 0.005. Notice that the ionization potential for the Ar(3p) state is 15.76 eV. The low-energy part $\sqrt{q^{2}_{x}+q^{2}_{y}}<0.1$ has been removed for better graphical display.
\begin{figure}
\begin{center}
    \includegraphics[width=0.494\columnwidth]{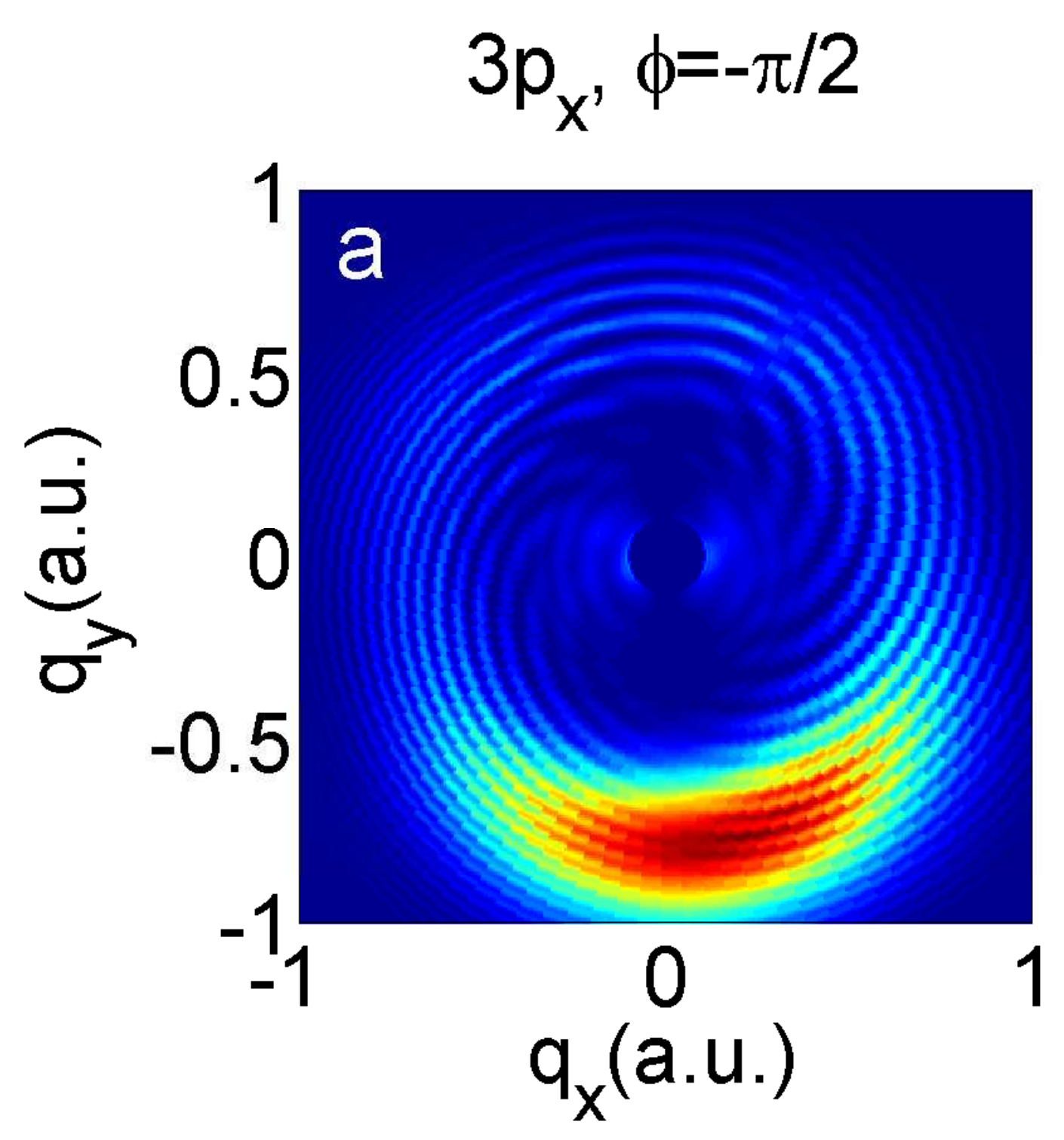}
    \includegraphics[width=0.494\columnwidth]{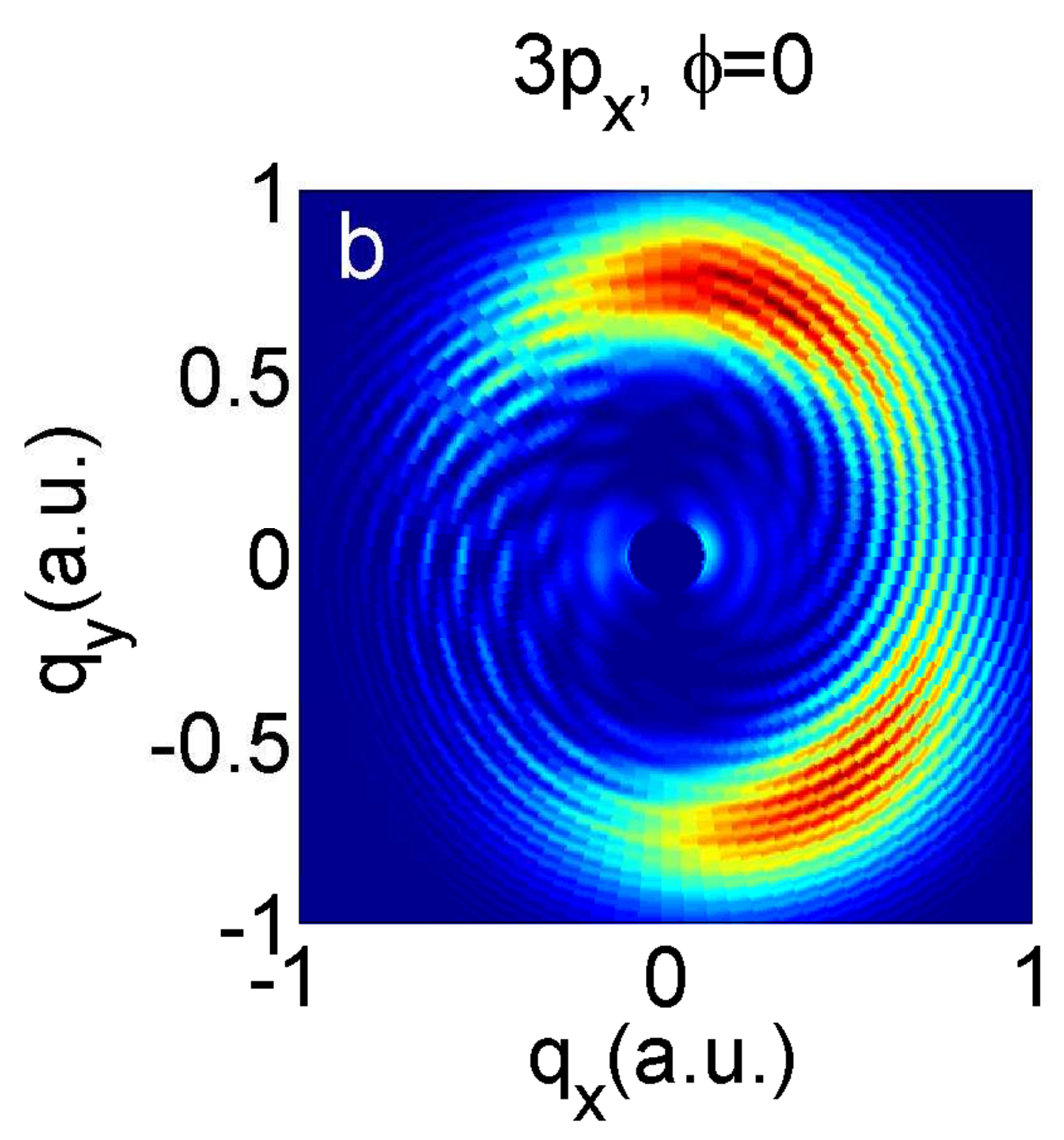}
    \includegraphics[width=0.494\columnwidth]{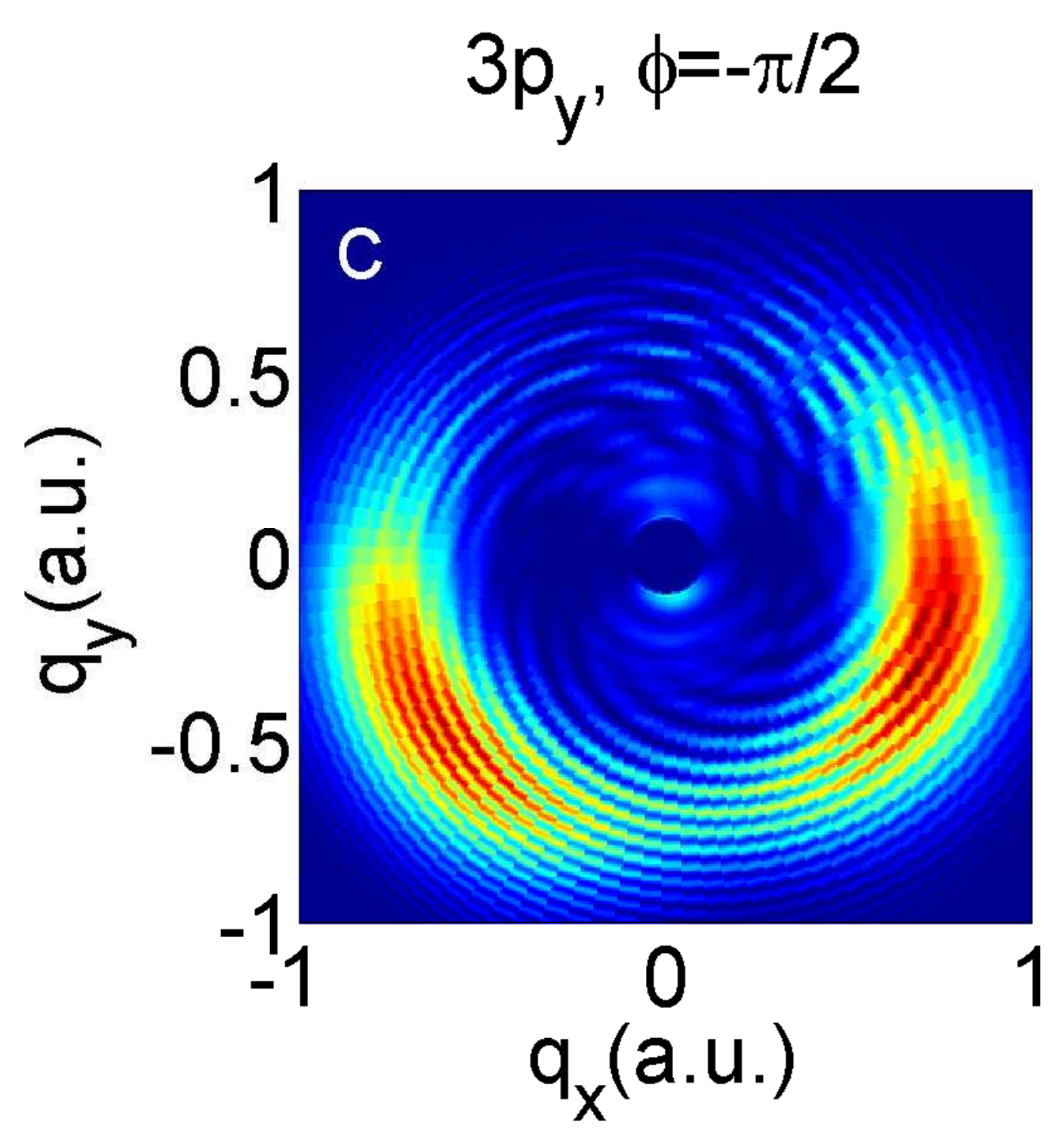}
    \includegraphics[width=0.494\columnwidth]{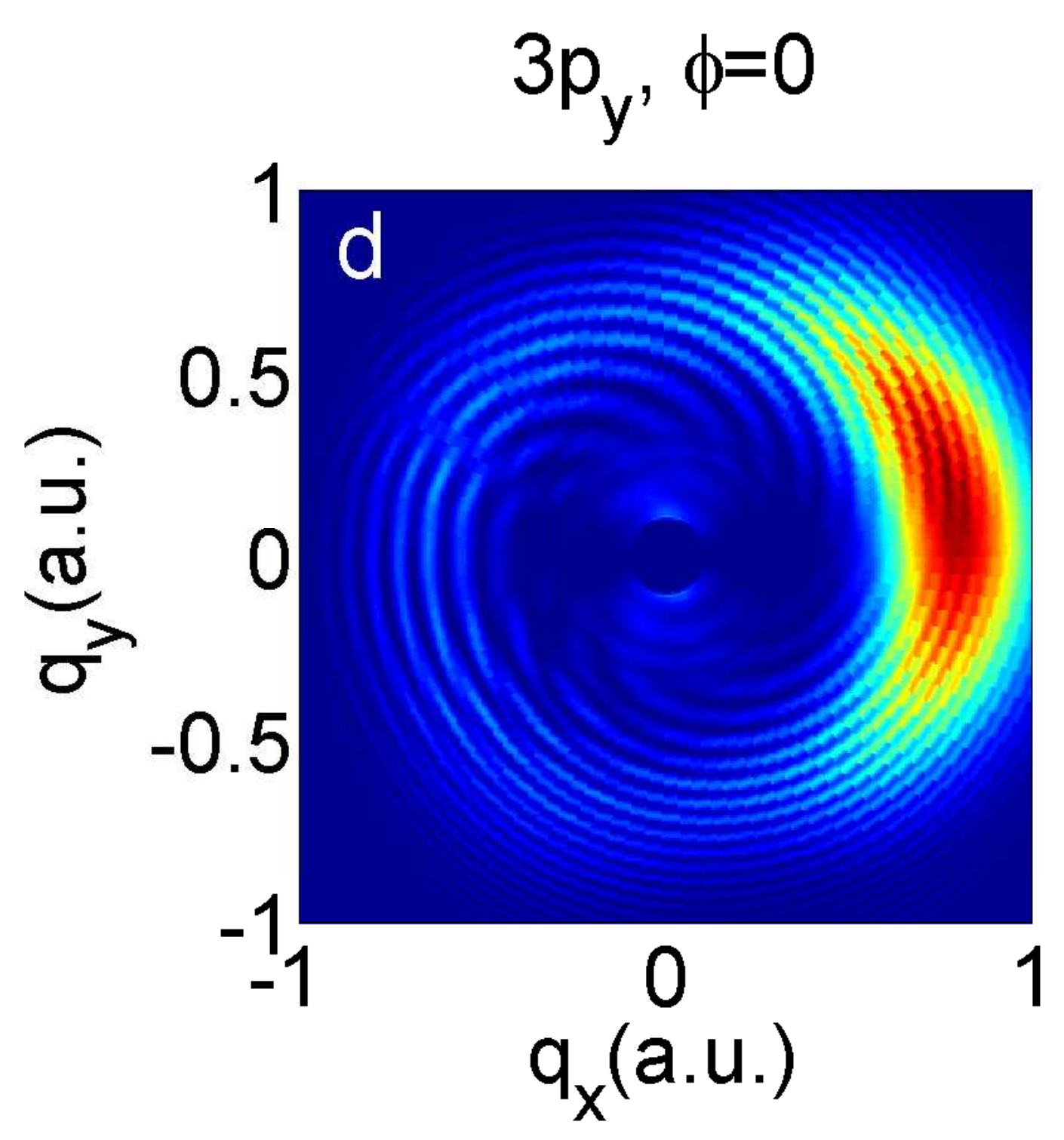}
    \caption{Momentum distributions from the single-active-electron TDSE calculation, for the (a, b) $3\text{p}_{x}$ and (c, d) $3\text{p}_{y}$ states of Ar, in the $xy$ polarization plane, for the following choice of laser parameters: Angular frequency $\omega=0.057$, corresponding to 800 nm, peak intensity $I=1.06 \times 10^{14}\text{W}/\text{cm}^{2}$, carrier envelope phase (a, c) $\varphi=-\pi/2$, (b, d) $\varphi=0$ and number of optical cycles $N=3$.}
\end{center}
\end{figure}
For the $3\text{p}_{x}$ state with $\phi=-\pi/2$, Fig. 1(a), and $3\text{p}_{y}$ state with $\phi=0$, Fig. 1(d), we observe a distribution with a single dominant peak, which, as we shall discuss in more detail below, is located near $\vec{q}\approx-\vec{A}(T/2)$. In the two other cases we observe a splitting of this peak into two nearly symmetric peaks. As we now show these main features are explained using the LG-SFA.
Figure 2 shows the momentum distributions obtained using the LG-SFA, in the plane of polarization, for the (a, b) $3\text{p}_{x}$ and (c, d) $3\text{p}_{y}$ initial states of Ar, at 800 nm light and peak intensity $I=1.06 \times 10^{14}\text{W}/\text{cm}^{2}$, CEP (a, c) $\varphi=-\pi/2$, (b, d) $\varphi=0$ and three optical cycles $N=3$.
\begin{figure}
\begin{center}
    \includegraphics[width=0.494\columnwidth]{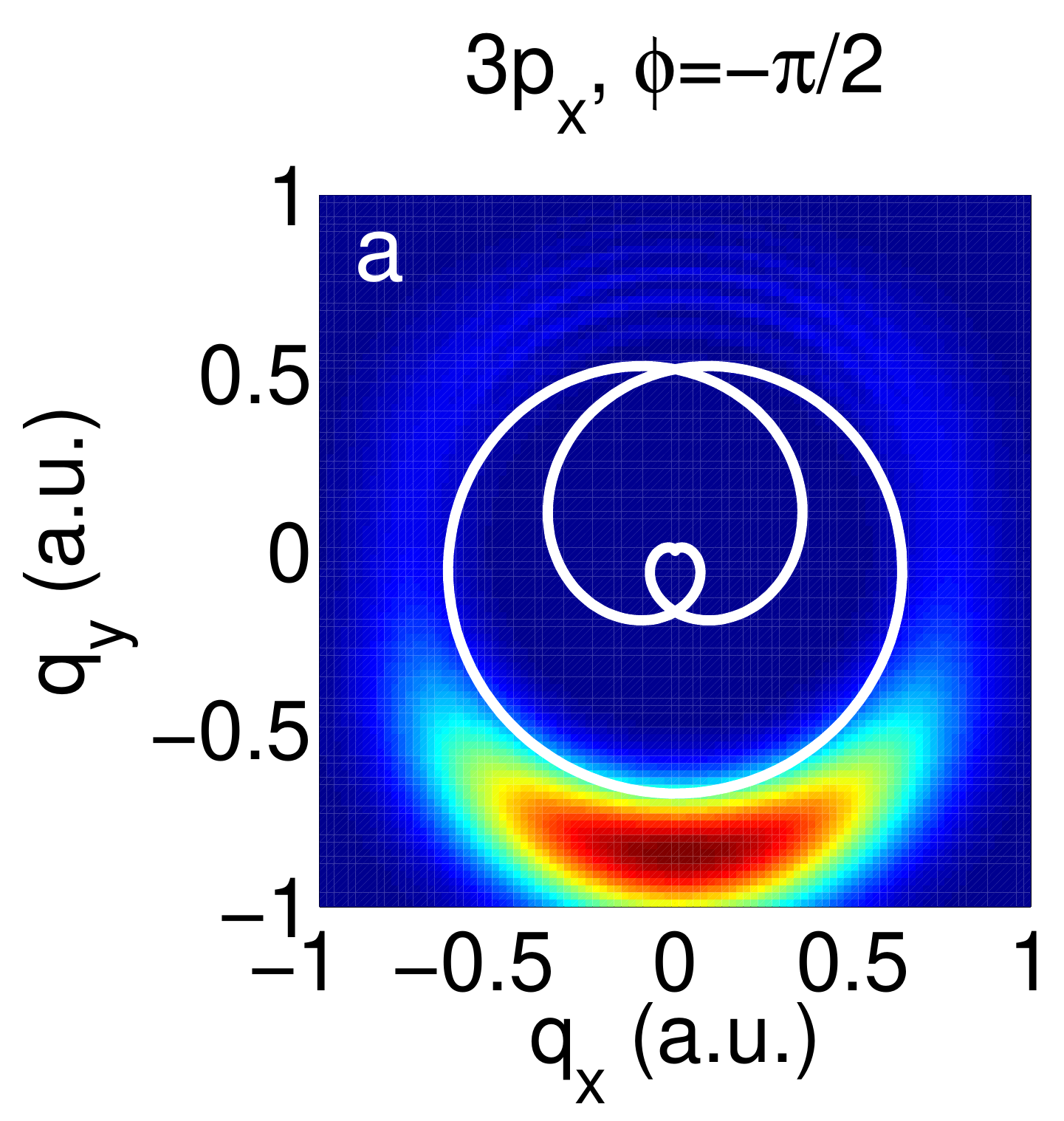}
    \includegraphics[width=0.494\columnwidth]{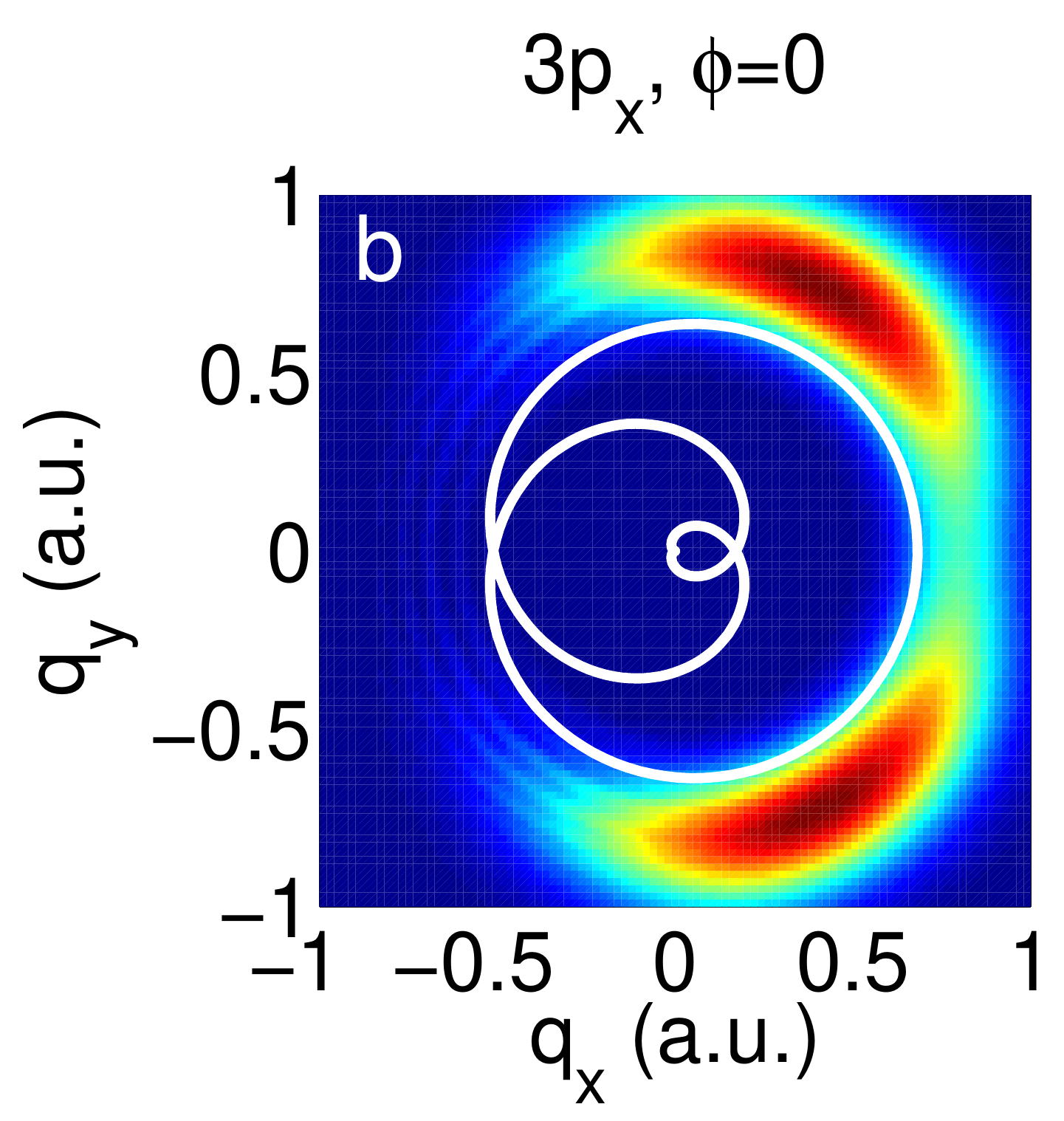}
    \includegraphics[width=0.494\columnwidth]{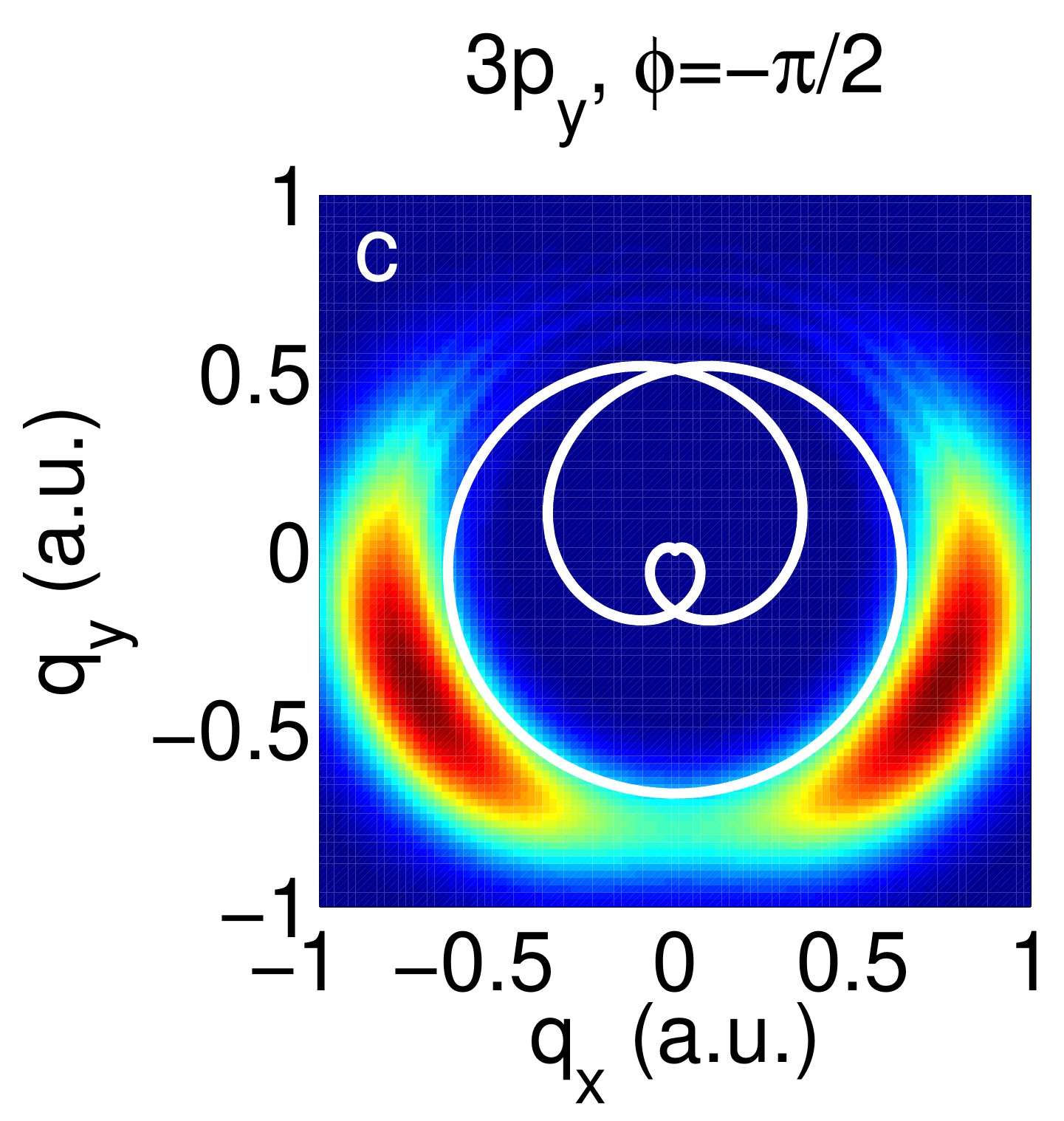}
    \includegraphics[width=0.494\columnwidth]{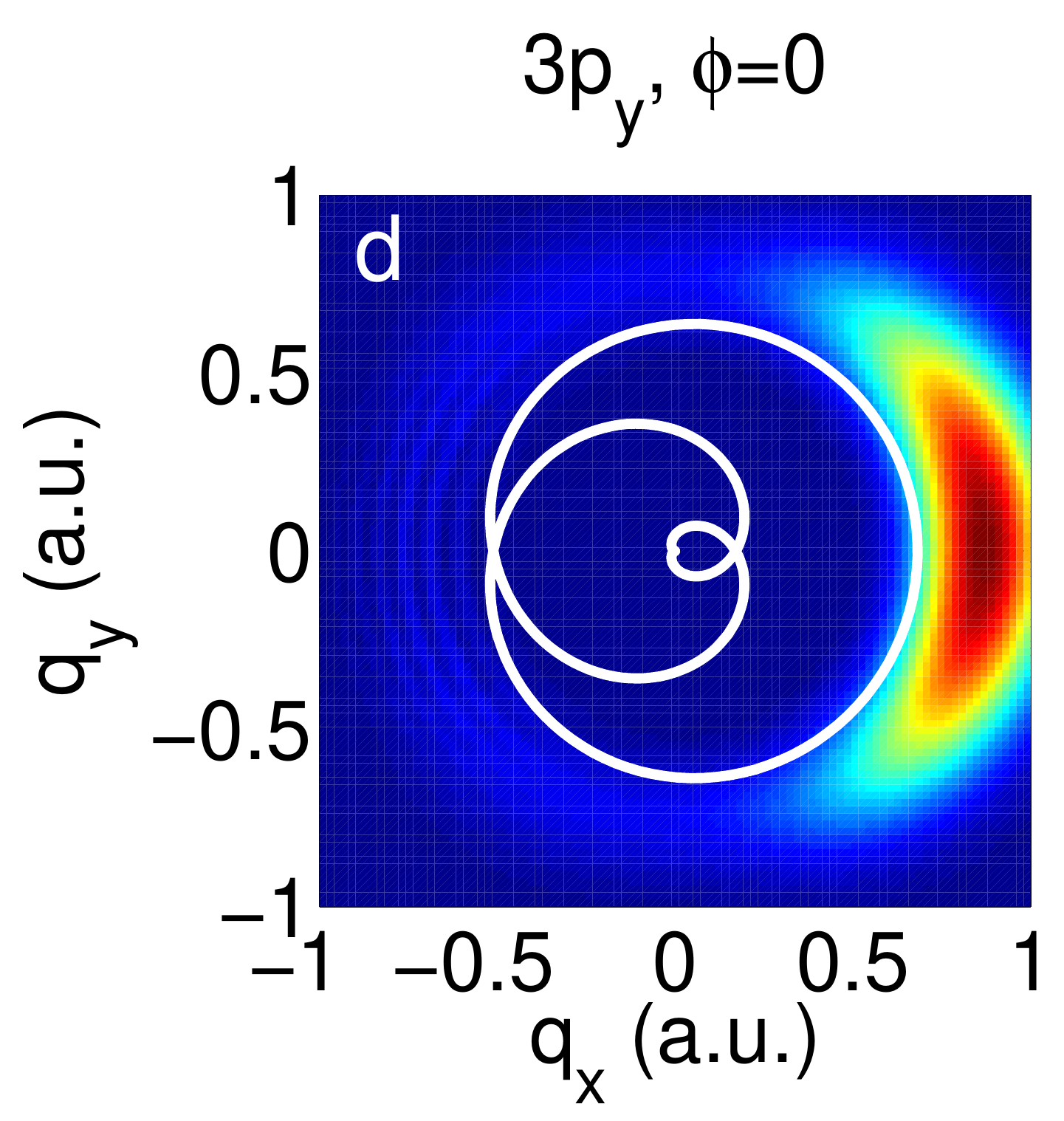}
    \caption{SFA momentum distributions, for the (a, b) $3\text{p}_{x}$ and (c, d) $3\text{p}_{y}$ states of Ar, in the $xy$ polarization plane, for the following choice of laser parameters: Angular frequency $\omega=0.057$, corresponding to 800 nm, peak intensity $I=1.06 \times 10^{14}\text{W}/\text{cm}^{2}$, CEP (a, c) $\varphi=-\pi/2$, (b, d) $\varphi=0$ and number of optical cycles $N=3$. The white curves show a parametric plot of $-\vec{A}(t)$ from Eq. (1). If no nodal planes are along the polarization at time $t=T/2$, the dominating peak in the distribution follows the simple formula $\vec{q}_{\text{final}}\simeq-\vec{A}(T/2)$ as seen in (a) and (d). If on the other hand the field peaks in the direction of the angular node (b), (c), the dominating peak in the distribution splits, simply because there is no charge density in that direction.}
\end{center}
\end{figure}
\begin{figure}
\begin{center}
    \includegraphics[width=0.494\columnwidth]{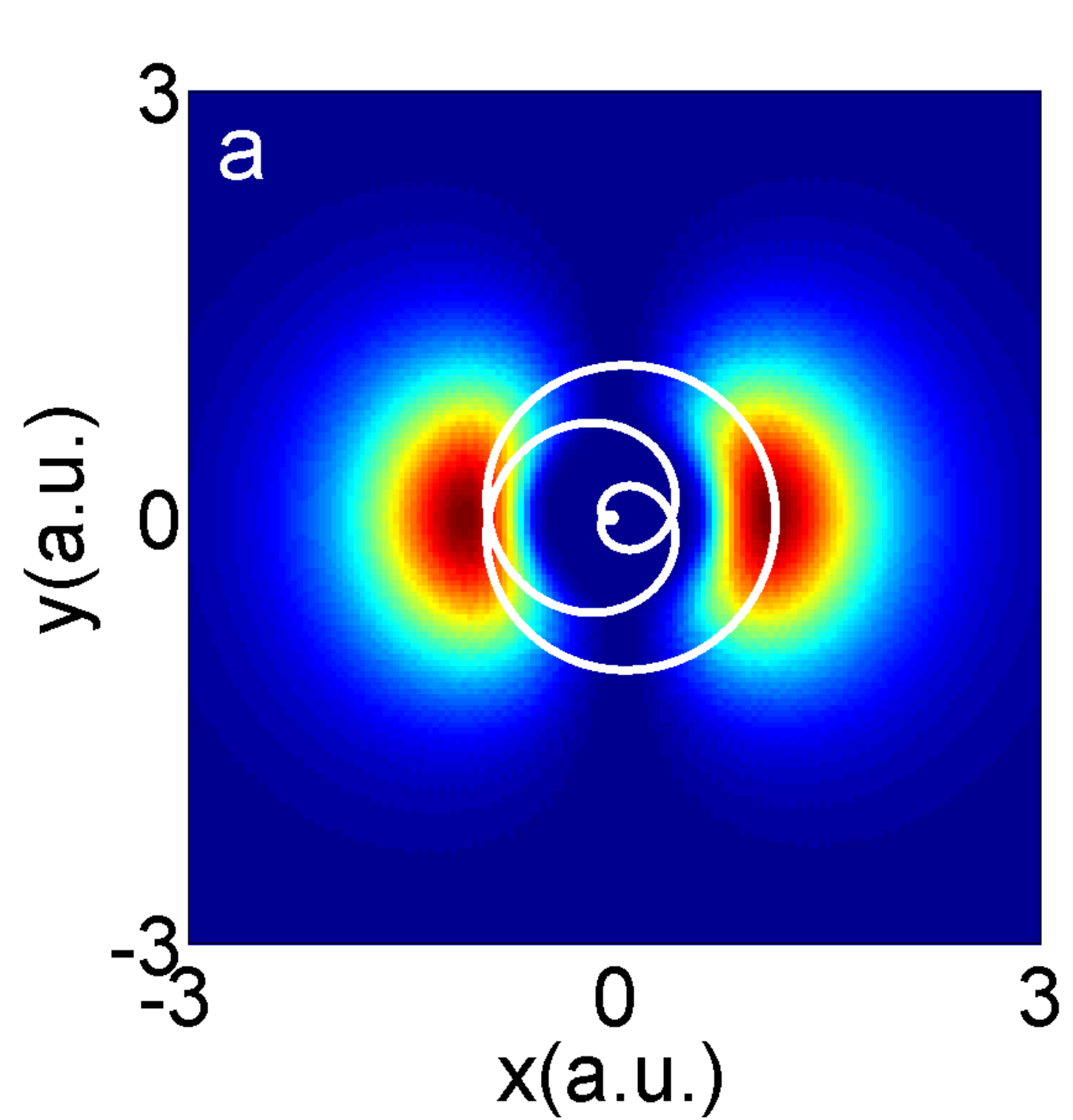}
    \includegraphics[width=0.494\columnwidth]{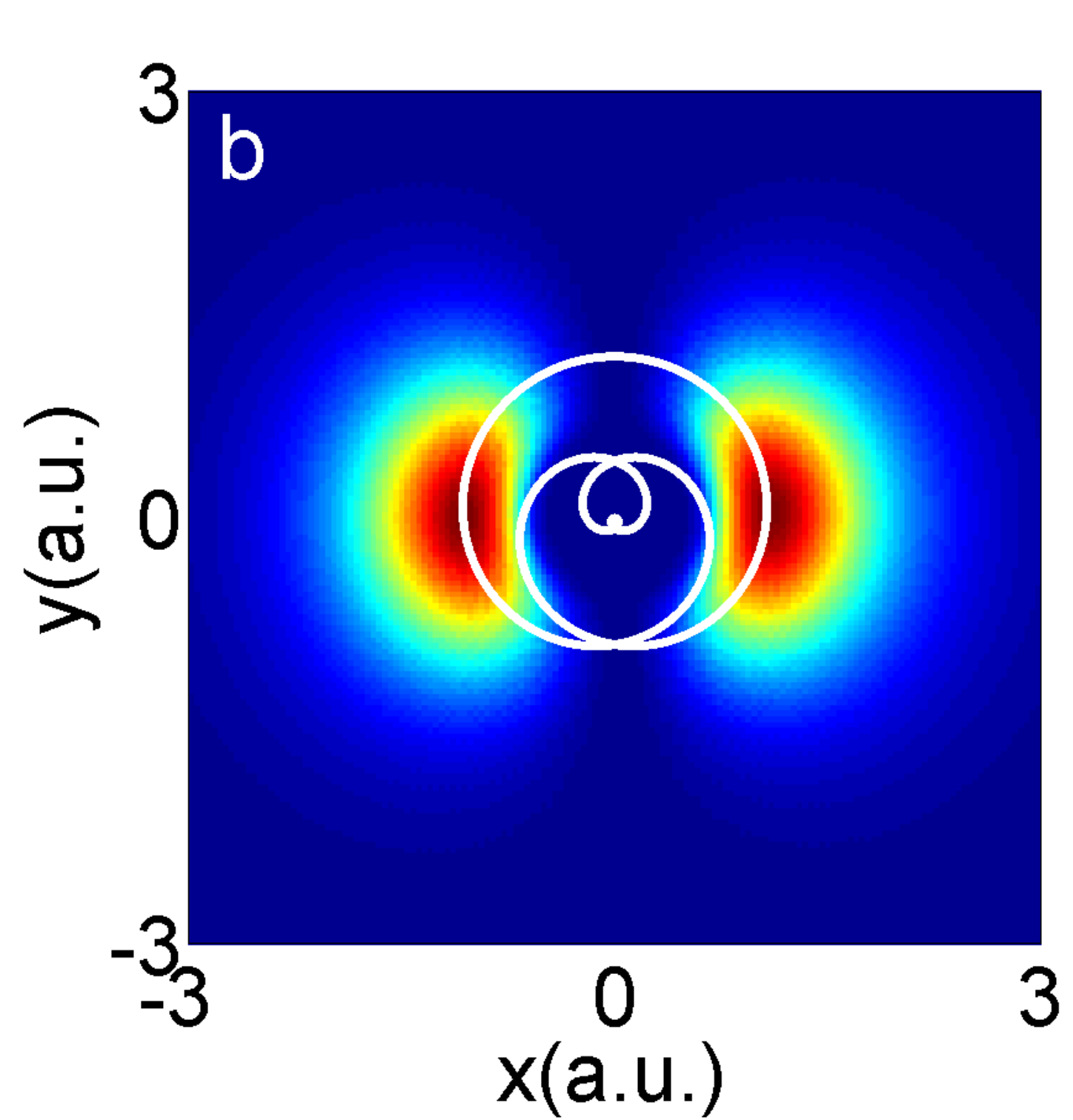}
    \caption{Electron density in the $3\text{p}_{x}$ state of Ar together with a parametric plot of the electric field, corresponding to the vector potential shown in Fig. 2, scaled by a factor. In (a), $\varphi=-\pi/2$, the electric field at the peak of the pulse $t=T/2$ points in a direction where the orbital density is largest. In (b), $\varphi=0$, the electric field at the peak of the pulse runs through the orbital node (no electron density).}
\end{center}
\end{figure}

The SFA results clearly reproduce the general structures of the TDSE distributions, except for an overall angular shift and a spiral like interference structure. The former overall angular shift is due partly to the Columbic interaction, partly to the fact that the laser field breaks the rotational invariance perpendicular to the polarization plane \cite{Mig5}. This feature has been observed experimentally in ionization of Helium by a few-cycle circularly polarized laser pulse \cite{Eckle2}. The spiral structure is connected to the interaction between outgoing low-energy electrons and the long-range Coulomb potential, equivalent to the radial fans seen in the linear case \cite{Rudenko, Maharjan, Morishita, Arbo, Mahmoud}.

It is well-known, in the case of a spherically symmetric initial state, that the LG-SFA momentum distribution reflects the shape of the vector potential \cite{Mig2}. This is easily explained by semiclassical theory \cite{Corkum}: Due to the exponential dependence of the rate on the field strength, ionization predominantly occurs at the times $t$ when the magnitude of the electric field is close to the maximum $|\vec{E}(T/2)|$. Assuming that the (classical) electron is born in the continuum with zero initial velocity and that it subsequently moves under the influence of the electric field only, we obtain $\vec{q}_{\text{final}}=-\vec{A}(t_{i})$ with $t_{i}$ the instant of ionization ($\vec{q}_{\text{final}}=-\vec{A}(T/2)$ dominates). However, in the case of a non-spherical initial state, this simple argument has to be extended to take into account the shape of the initial state, in particular the angular nodal structure. For example, it may happen that the field is in a nodal plane of the initial state at $t=T/2$ and hence ionization will be suppressed. In other words, it is in this more general case, the shape of the vector potential, electric field and initial state that determines the overall structure of the momentum distribution. The results in Figs. 1 and 2 clearly illustrate this combined effect. Let us start by analyzing the results for the $3\text{p}_{x}$ state and let us concentrate on the LG-SFA results to make the discussion as simple as possible.
Figure 2(a) shows the SFA momentum distribution in the plane of polarization for $\phi=-\pi/2$ and $3\text{p}_{x}$ initial state. We observe a distribution with a single peak located near $-\vec{A}(T/2)$, in good agreement with the semiclassical theory. In the case of $\varphi=-\pi/2$, $\vec{E}(T/2)=E_{x}(T/2)\hat{x}$ points in the direction of maximum electron density (see Fig. 3(a)), while both $\vec{E}$ and $3\text{p}_{x}$ are symmetric with respect to $(x,y)\rightarrow(x,-y)$. This leads to a single peak situated at $\vec{q}\sim-\vec{A}(T/2)\propto-\hat{y}$ and symmetric with respect to $(x,y)\rightarrow(-x,y)$ in good agreement with the calculated distribution. Figure 2(b) shows the SFA momentum distribution, in the plane of polarization, for $\phi=0$ and a $3\text{p}_{x}$ initial state. We observe two peaks, symmetric with respect to $(q_{x},q_{y})\rightarrow(q_{x},-q_{y})$, in contrast to the simplest single peak semiclassical prediction. In the case of $\varphi=0$, $\vec{E}(T/2)=E_{y}(T/2)\hat{y}$ lies in the nodal plane of $3\text{p}_{x}$ (see Fig. 3(b)), i.e., there is no electron density to ionize at $t=T/2$ and hence we do not have ionization at field maximum. When $|t-T/2|$ increases from zero the electric field begins to point in directions with non-zero electronic density. Thus, the single peak from before splits into two symmetric peaks, due to the fact that $\vec{E}$ and $3\text{p}_{x}$ are symmetric with respect to $(x,y)\rightarrow(-x,y)$, again in good agreement with the obtained distribution. The precise location of the peaks is determined by a competition between the rate arising from the electric field and the density profile of the initial state. The results shown in Fig. 2(c) and Fig. 2(d) are explained using similar reasoning. Notice, that the signature of the nodal plane in the momentum distribution is advanced by $\pi/2$ compared to the orbital angular nodal structure, reflecting the $\pi/2$ phase between the electric field and the vector potential.

The above discussion of Figs. 1, 2 and 3 clearly shows that the nodal structure of the initial orbital is mapped uniquely to the momentum distribution when the laser polarization plane is perpendicular to the nodal plane. This effect is even more visible in the case of a slightly longer femtosecond pulse, where the asymmetry caused by the CEP of the pulse is smaller. Figure 4 shows the SFA reduced momentum distribution in the plane of polarization $\partial^{2}P/\partial q_{x} \partial q_{y}=\int(\partial^{3}P/\partial q_{x} \partial q_{y} \partial q_{z} )dq_{z}$ for H(1s), H($2\text{p}_{x}$), H($2\text{p}_{y}$) and H($3\text{d}_{xy}$) states obtained using the same parameters as in Figs. 1 and 2 except that the number of optical cycles was increased to $N=20$ and $I=1.0 \times 10^{14}\text{W}/\text{cm}^{2}$. The use of the LG-SFA is for computational convenience and is justified by the results presented on Ar. The binding energy of the H($2\text{p}_{x}$), H($2\text{p}_{y}$) and H($3\text{d}_{xy}$) have been artificial modified to $-0.5$, by modifying $\kappa=\sqrt{2I_{p}}$, in order to compare with the H(1s) result. Figure 4 clearly shows that the angular nodal structure is mapped uniquely to the momentum distribution. This is particular clear when compared with Fig. 5, which shows the asymptotic densities for the 1s, $2\text{p}_{x}$, $2\text{p}_{y}$ and $3\text{d}_{xy}$ states. The electric field is almost constant during the dominant cycles of the field, i.e., there is no competition between the rate arising from the electric field, and the density profile. In the case of H(1s) we observe a completely symmetric distribution with well-resolved above-threshold ionization peaks, while ionization is strongly suppressed along the $x$-axis ($y$-axis) in the case of H($2\text{p}_{x}$) (H($2\text{p}_{y}$)) and along the $x$-axis and $y$-axis in the case of H($3\text{d}_{xy}$), reflecting the angular nodal structures in Fig. 5. Ionization is suppressed along the $x$-axis ($y$-axis) in the case of H($2\text{p}_{x}$) (H($2\text{p}_{y}$)), due to the $yz$ ($xz$) nodal plane which is shifted $90^{\circ}$ by the vector potential. There are two angular nodal planes for H($3\text{d}_{xy}$), namely the $xz$ and $yz$ planes, suppressing ionization along the $y$-axis and $x$-axis.
\begin{figure}
\begin{center}
    \includegraphics[width=0.494\columnwidth]{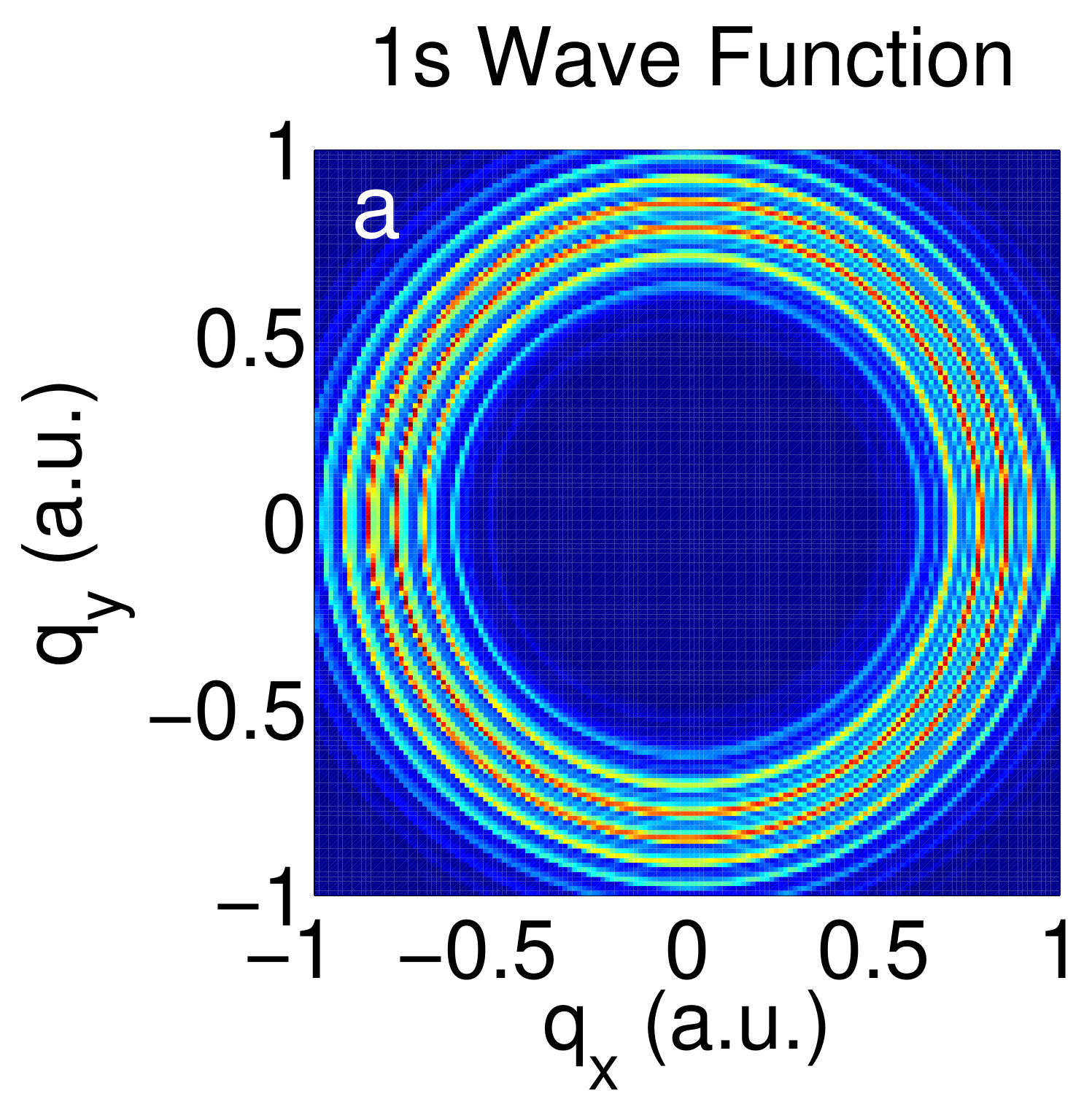}
    \includegraphics[width=0.494\columnwidth]{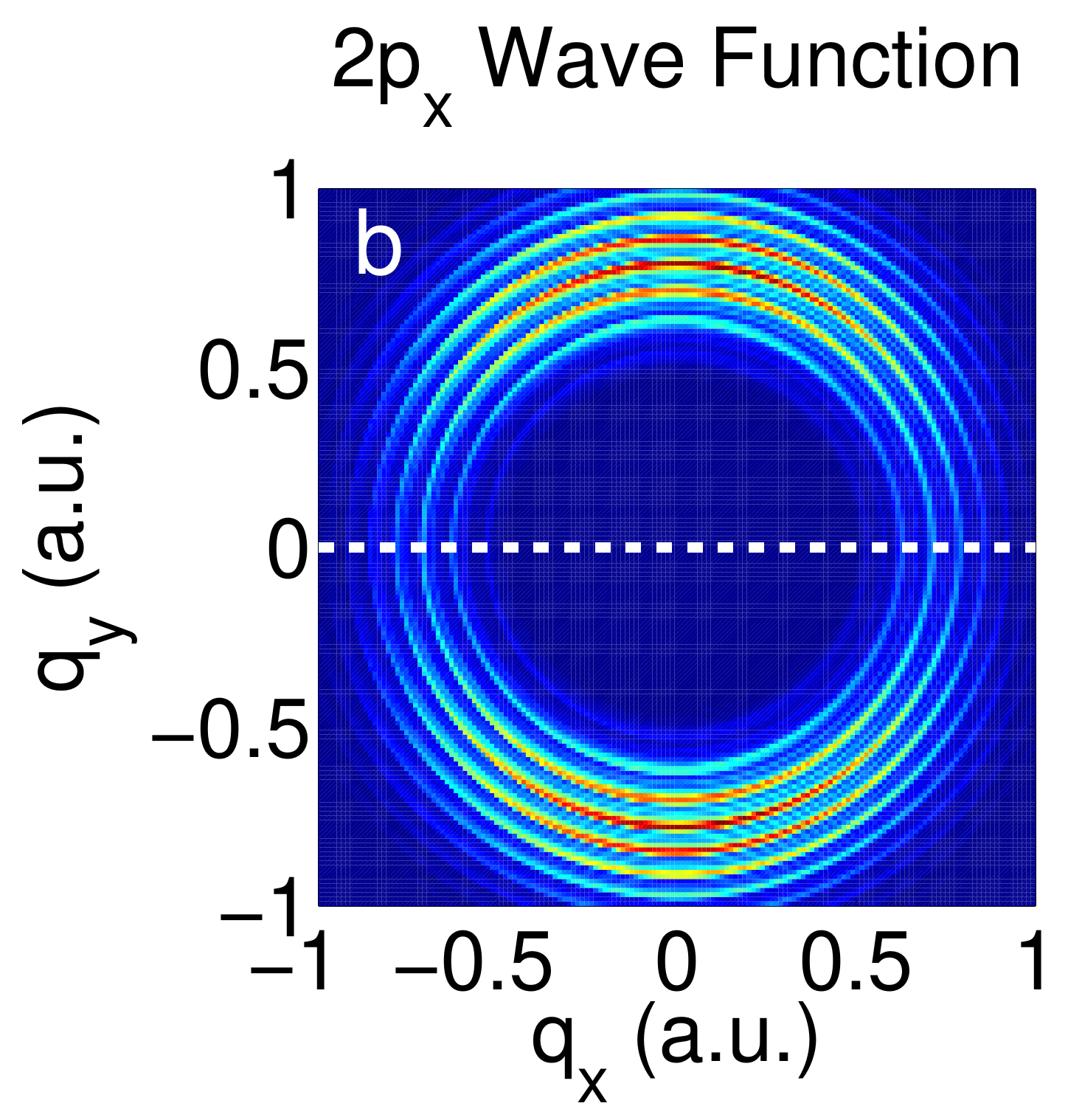}
    \includegraphics[width=0.494\columnwidth]{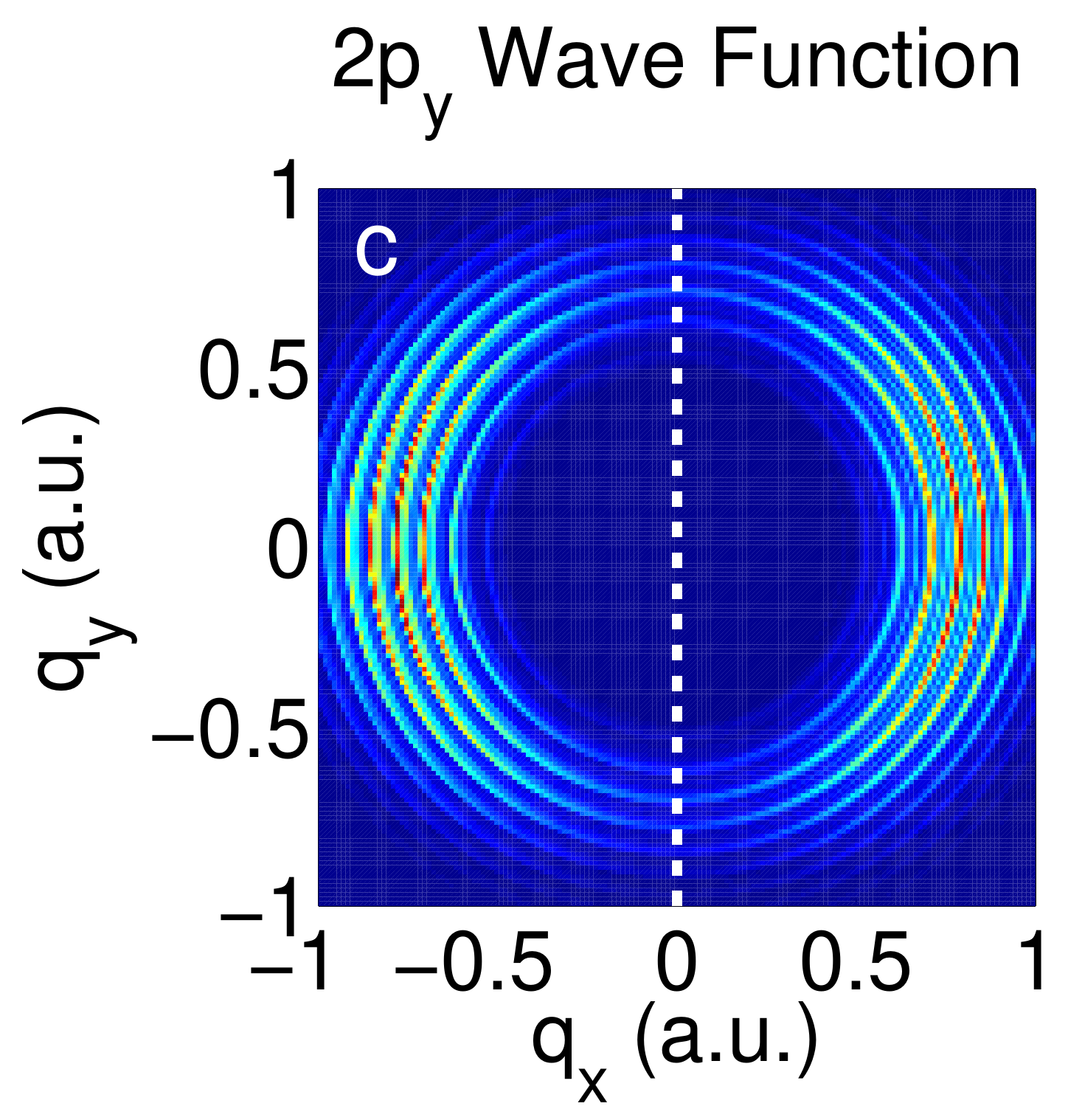}
    \includegraphics[width=0.494\columnwidth]{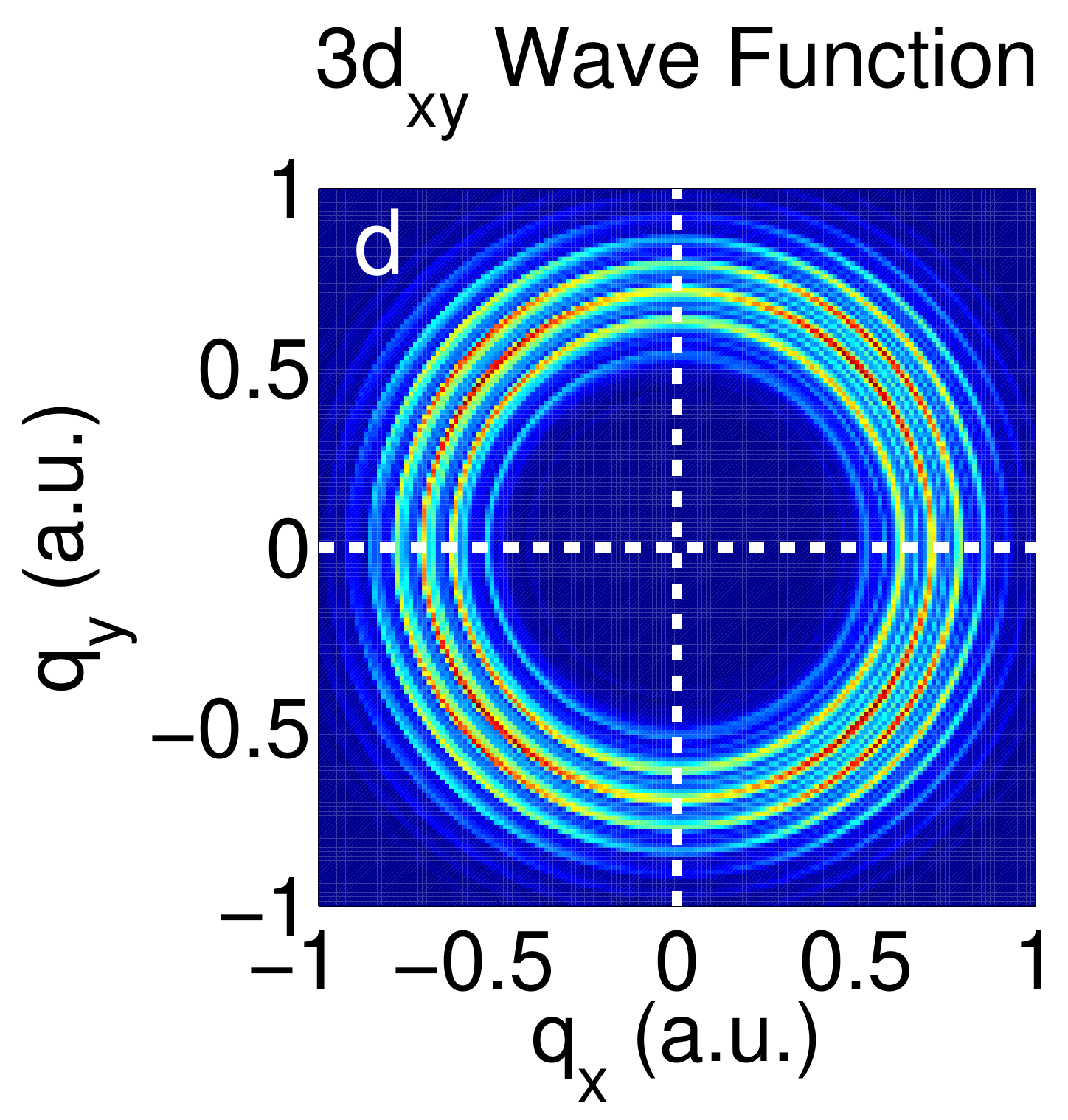}
    \caption{SFA momentum distributions $\partial^{2}P/\partial q_{x} \partial q_{y}=\int(\partial^{3}P/\partial q_{x} \partial q_{y} \partial q_{z}) dq_{z}$ for (a) 1s, (b) $2\text{p}_{x}$, (c) $2\text{p}_{y}$ and (d) $3\text{d}_{xy}$ obtained using the following laser parameters: Angular frequency $\omega=0.057$, corresponding to 800 nm, peak intensity $I=1.0 \times 10^{14}\text{W}/\text{cm}^{2}$, carrier envelope phase $\varphi=0$ and number of optical cycles $N=20$. The binding energy of the $2\text{p}_{x}$, $2\text{p}_{y}$ and $3\text{d}_{xy}$ states are modified to the ground state energy of H. The white dashed lines show regions where ionization is suppressed.}
\end{center}
\end{figure}

\begin{figure}
\begin{center}
    \includegraphics[width=0.494\columnwidth]{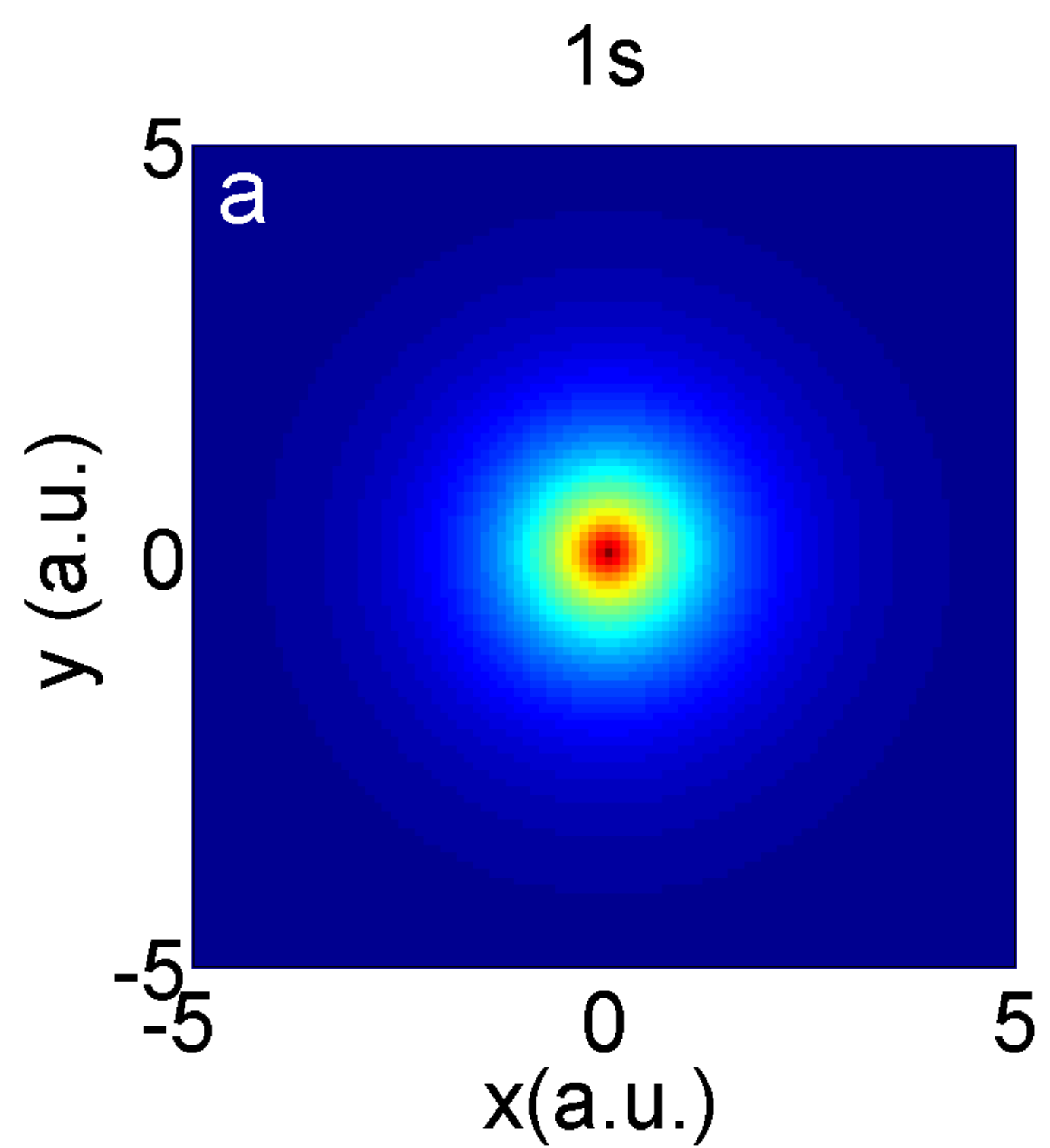}
    \includegraphics[width=0.494\columnwidth]{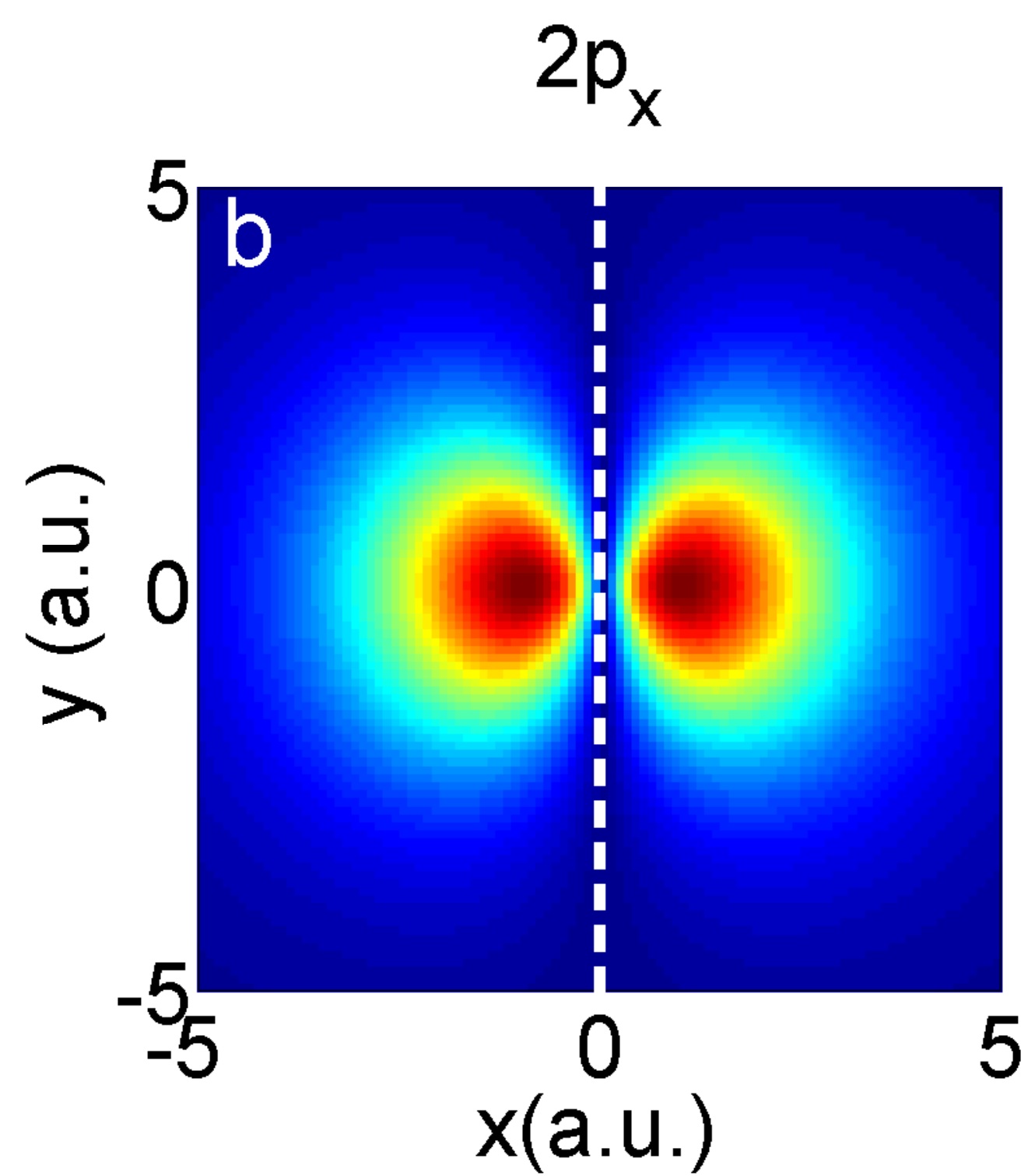}
    \includegraphics[width=0.494\columnwidth]{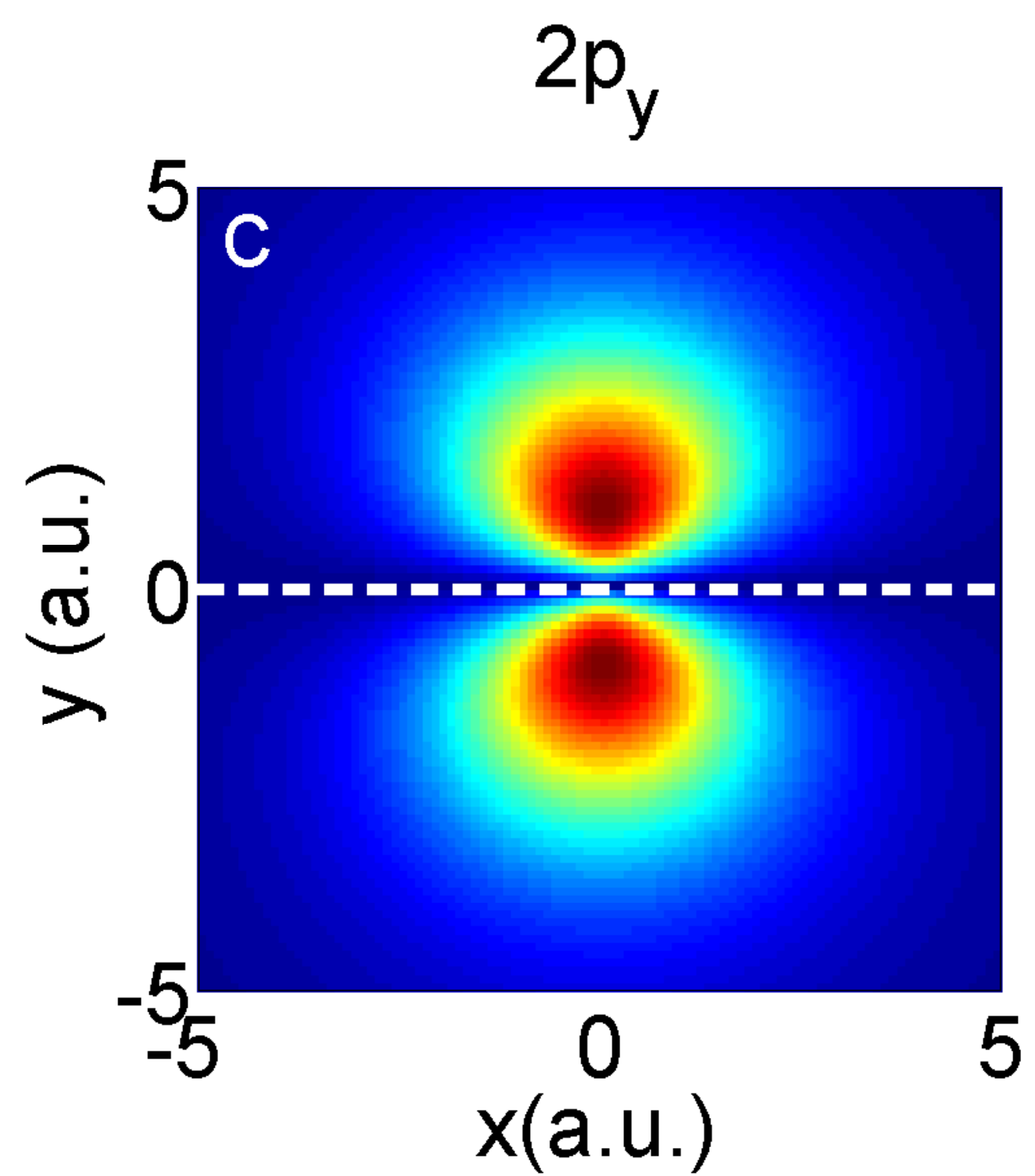}
    \includegraphics[width=0.494\columnwidth]{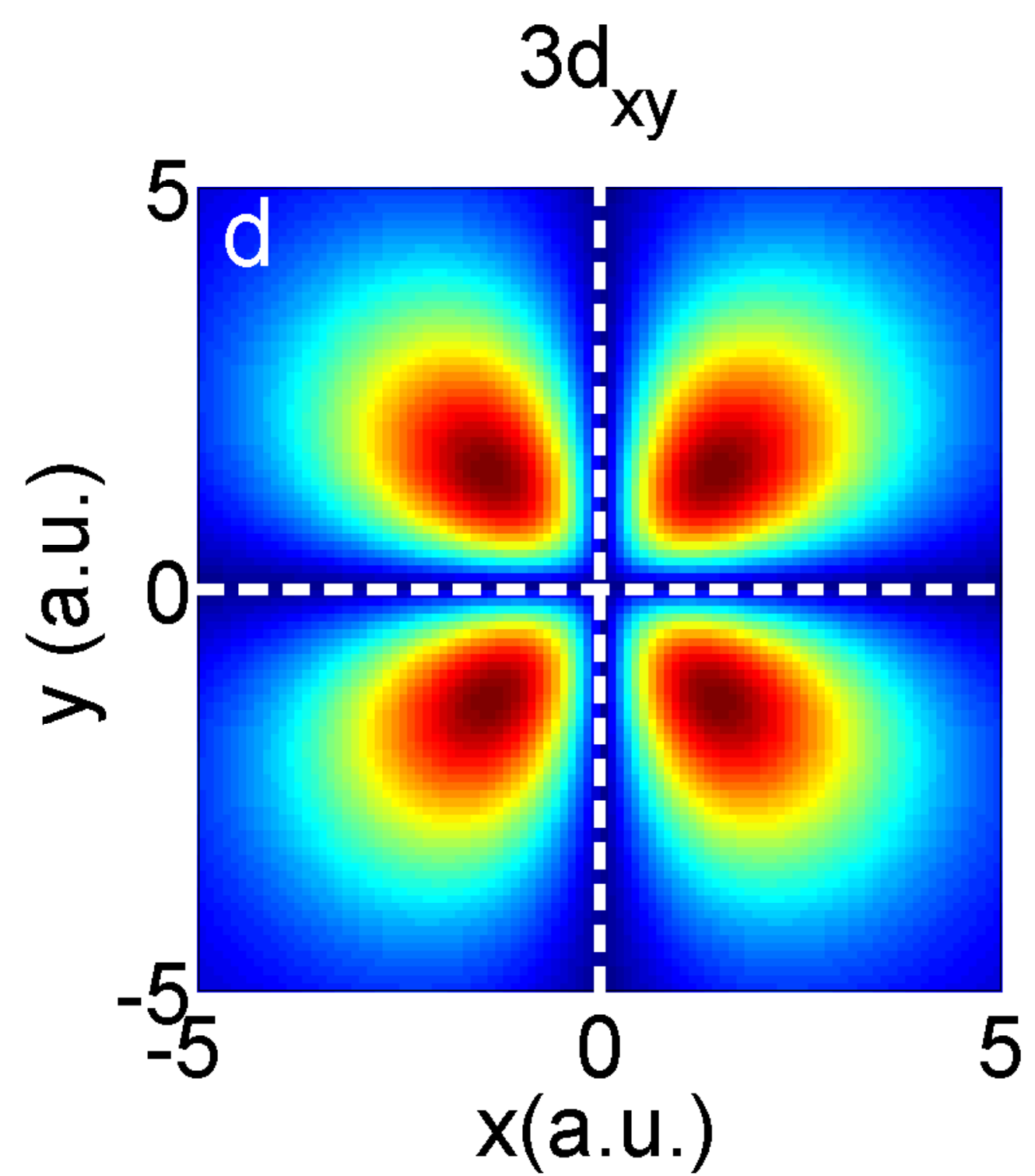}
    \caption{The asymptotic density $|\sum_{lm}e^{-r}Y_{lm}|^{2}$ for (a) H(1s), (b) H($2\text{p}_{x}$), (c) H($2\text{p}_{y}$) and (d) H($3\text{d}_{xy}$. The white dashed lines show the angular nodal planes.}
\end{center}
\end{figure}

\subsection{Carrier-envelope phase effects in circularly polarized fields}
In Figs. 1 and 2, we note that changing the CEP from $\varphi=-\pi/2$ to $\varphi=0$ does not correspond to a counterclockwise rotation of angle $\pi/2$, in contrary to what one might expect: CEP effects for atoms and linear molecules aligned along the propagation axis of the applied circularly polarized field were characterized in terms of simple rotations of the total system around the propagation axis \cite{Mig1}. This characterization, however, only holds when the initial state is invariant with respect to rotations around the propagation direction or when the system initially is described by a uniform incoherent mixture of magnetic substates. The latter point was not discussed explicitly in \cite{Mig1} and we use the opportunity to do so now. Note that CEP effects have been described generally in a Floquet-like approach \cite{Esry, Esry2}.
\begin{figure}
\begin{center}
    \includegraphics[width=0.494\columnwidth]{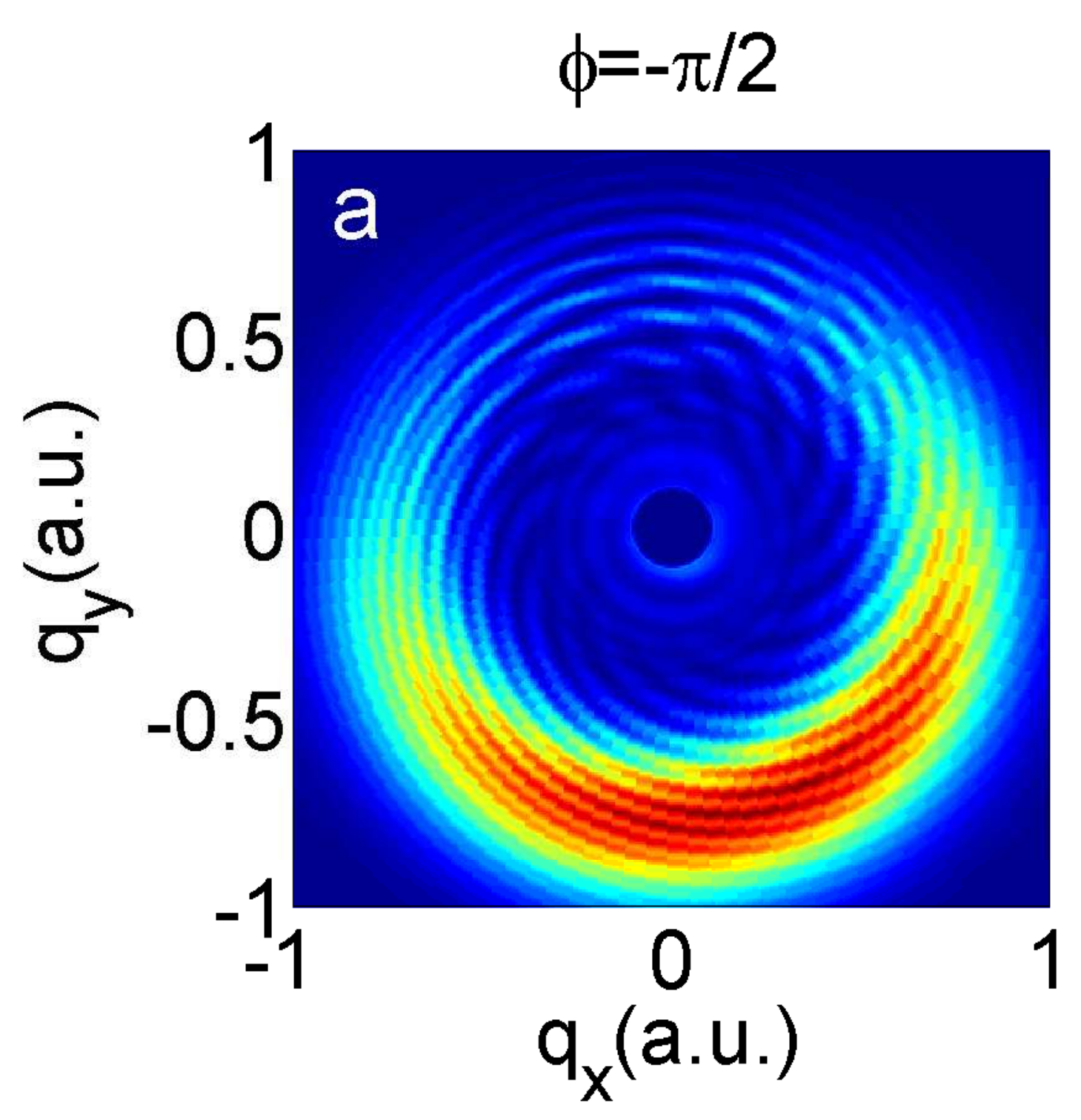}
    \includegraphics[width=0.494\columnwidth]{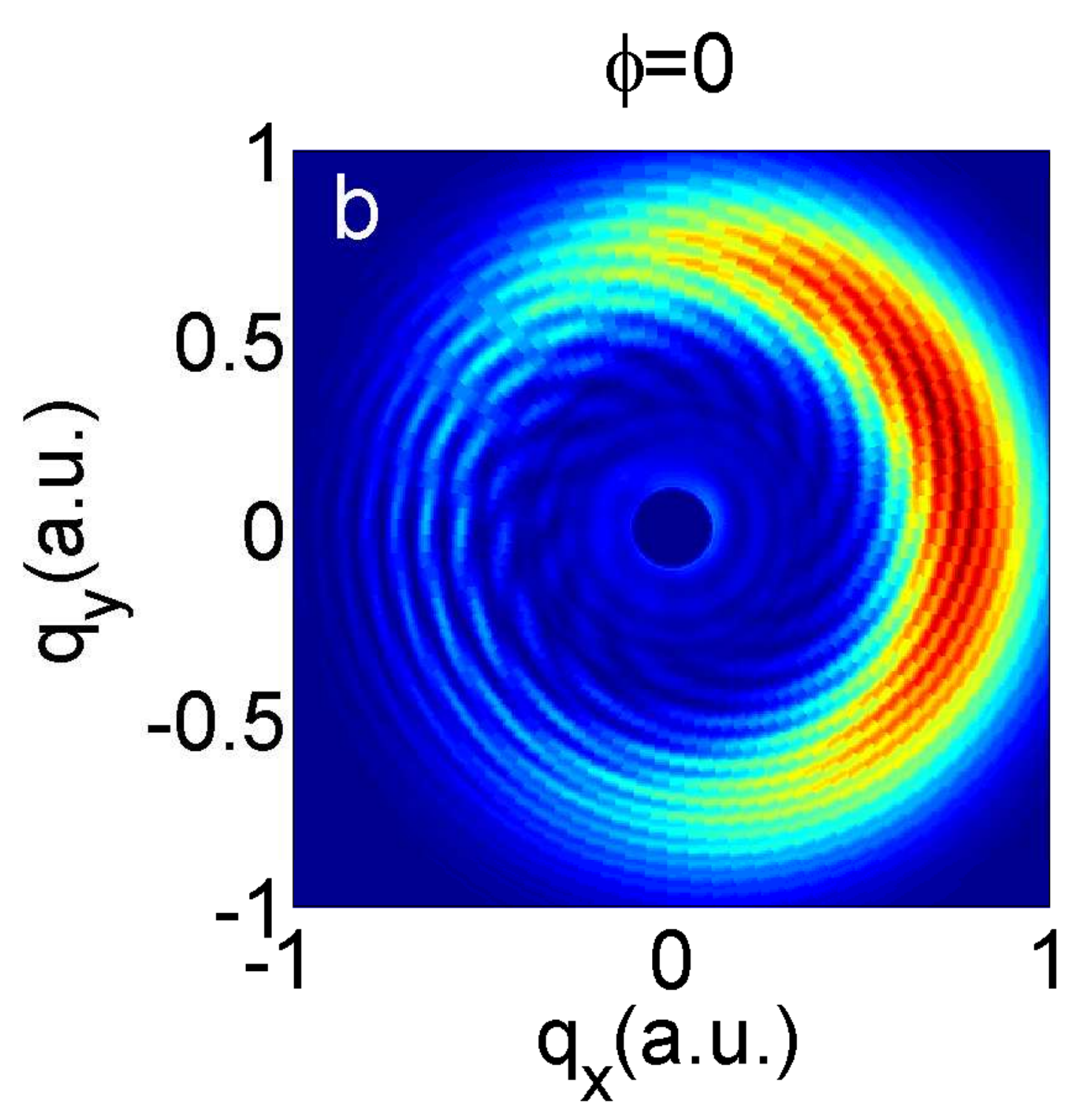}
    \caption{Ensemble average of the SAE, TDSE momentum distributions in the $xy$ polarization plane for Argon initially in a uniform incoherent mixture of $3\text{p}_{x}$ and $3\text{p}_{y}$ states. The laser parameters are: Angular frequency $\omega=0.057$, corresponding to 800 nm, peak intensity $I=1.06 \times 10^{14}\text{W}/\text{cm}^{2}$, carrier envelope phase (a) $\phi=-\pi/2$, (b) $\phi=0$ and number of optical cycles $N=3$.}
\end{center}
\end{figure}

Consider an atom or a linear molecule, described within the SAE, interacting with a
laser pulse described by the vector potential $\vec{A}(t)$ defined in Eq. (1). We assume that the field-free Hamiltonian $H_{0}$ is invariant under rotations around the $z$-axis. Furthermore, we assume that the probability for a specific magnetic quantum number $P_{M}$ is uniform. The ensemble average of $\partial^{3}P/\partial q_{x} \partial q_{y} \partial q_{z}$ is defined as \cite{Sakurai}
\begin{equation}
\left[\frac{\partial^{3}P}{\partial q_{x} \partial q_{y} \partial q_{z}}(\vec{q},\phi)\right]=\text{Tr}(\rho(\phi;t) \hat{P}_{\vec{q}}),
\end{equation}
where $t$ is any time after the pulse, $\rho(\phi;t)$ is the density matrix of the total system, $\phi$ the CEP and $\hat{P}_{\vec{q}}=|\Psi^{-}_{\vec{q}}\rangle\langle\Psi^{-}_{\vec{q}}|$ projects on the exact scattering states $|\Psi^{-}_{\vec{q}}\rangle$ with asymptotic momentum $\vec{q}$. The density matrix is given by
\begin{equation}
\rho(\phi;t)=\sum^{J}_{M=-J}P_{M}|\Psi_{nJM}(\phi;t)\rangle\langle\Psi_{nJM}(\phi;t)|.
\end{equation}
Here $P_{M}=\frac{1}{2J+1}$ and $|\Psi_{nJM}(\phi;t)\rangle=U(\phi;t,0)|nJM\rangle$, where $U$ is the time-evolution operator for the total system and $|nJM\rangle$ the field-free initial state with $J$ denoting the total angular momentum of the system, $M$ the corresponding magnetic quantum number and $n$ the remaining quantum numbers. We note, in passing, that the Hamiltonian and hence the time-evolution operator have a parametric dependence on $\phi$ through the interaction with the external field.
We evaluate the ensemble average Eq. (4) using the position-eigenstate basis and thereby obtain
\begin{equation}
\left[\frac{\partial^{3}P}{\partial q_{x} \partial q_{y} \partial q_{z}}(\vec{q},\phi)\right]=\frac{1}{2J+1}\sum^{J}_{M=-J}\left|\int d\vec{r}(\Psi^{-}_{\vec{q}})^{\ast}\Psi_{nJM}(\phi;t)\right|^{2}.
\end{equation}
Using the relation \cite{Zare}
\begin{equation}
\int d\Omega D^{(J)}_{M'M}(\alpha,\beta,\gamma)D^{(J)\ast}_{M''M}(\alpha,\beta,\gamma)=\frac{8\pi^{2}}{2J+1}\delta_{M'M''},
\end{equation}
where $D^{(J)}_{M'M}(\alpha,\beta,\gamma)$ are the Wigner rotation functions, $(\alpha,\beta,\gamma)$ the Euler angles and $d\Omega=\sin(\beta)d\beta d\alpha d\gamma$, Eq. (6) can also be expressed as
\begin{equation}
\left[\frac{\partial^{3}P}{\partial q_{x} \partial q_{y} \partial q_{z}}(\vec{q},\phi)\right]=\frac{1}{8\pi^{2}}\int d\Omega\left|\int d\vec{r}(\Psi^{-}_{\vec{q}})^{\ast}\Psi_{nJM'}(\Omega,\phi;t)\right|^{2}.
\end{equation}
Here $\Psi_{nJM'}(\Omega,\phi;t)$ is a solution to the TDSE corresponding to the rotated initial state $D(\Omega)|nJM'\rangle$, with $D$ the rotation operator and $\Omega=(\alpha,\beta,\gamma)$.
However $\Psi_{nJM'}(\Omega,\phi;t)=\exp(-i\hat{J}_{z}\phi)\Psi_{nJM'}(\Omega',\phi=0;t)$ with $\Omega'=(\alpha-\phi,\beta,\gamma)$ \cite{Mig1}. Thus
\begin{align}
\left[\frac{\partial^{3}P}{\partial q_{x} \partial q_{y} \partial q_{z}}(\vec{q},\phi)\right]=\frac{1}{8\pi^{2}}\int d\Omega\left|\int d\vec{r}\ (\Psi^{-}_{\vec{q}})^{*}\exp(-i\hat{J}_{z}\phi)\Psi_{nJM'}(\Omega',\phi=0;t)\right|^{2}.\
\end{align}
The rotation of a scattering wave function with asymptotic momentum $\vec{q}$ can be accomplished just by rotating the asymptotic momentum, i.e.,
$\exp(i\hat{J}_{z}\phi)\Psi^{-}_{\vec{q}}=\Psi^{-}_{\vec{q'}}$, where $\vec{q'}=\mathbf{R}_{z}(-\phi)\vec{q}$, with $\mathbf{R}_{z}$ the $3\times3$ orthogonal matrix which generates counterclockwise rotations around the $z$-axis \cite{Zare}. This means that
\begin{align}
\left[\frac{\partial^{3}P}{\partial q_{x} \partial q_{y} \partial q_{z}}(\vec{q},\phi)\right]&=\frac{1}{8\pi^{2}}\int^{2\pi}_{0}\int^{2\pi}_{0}\int^{\pi}_{0} \sin(\beta)d\beta d\alpha d\gamma\left|\int d\vec{r}\ (\Psi^{-}_{\vec{q'}})^{*}\Psi_{nJM'}(\Omega',\phi=0;t)\right|^{2}.
\end{align}
Finally this expression can be rewritten as
\begin{align}
\left[\frac{\partial^{3}P}{\partial q_{x} \partial q_{y} \partial q_{z}}(\vec{q},\phi)\right]&=\frac{1}{8\pi^{2}}\int^{2\pi}_{0}\int^{2\pi-\phi}_{-\phi}\int^{\pi}_{0} \sin(\beta)d\beta d\alpha d\gamma\left|\int d\vec{r}\ (\Psi^{-}_{\vec{q'}})^{*}\Psi_{nJM'}(\Omega,\phi=0;t)\right|^{2}\nonumber\\
&=\left[\frac{\partial^{3}P}{\partial q_{x} \partial q_{y} \partial q_{z}}(\vec{q'},\phi=0)\right],
\end{align}
due to the uniformity of the distribution function over the orientations, $G(\Omega)=1/(8\pi^{2})$.
Thus a change in the CEP from $\phi=0$ to $\phi=\phi'$ corresponds to a rotation of the system around the $z$-axis by $\phi'$.
This is illustrated in Fig. 6, which shows the ensemble average of the exact momentum distribution, in the $xy$ polarization plane, for Ar atoms initially in an incoherent mixture of  $3\text{p}_{x}$ or $3\text{p}_{y}$ states, for two different values of the CEP $\phi=-\pi/2$ and $\phi=0$. Notice, that the validity of Eq. (11) does not require the SAE or the dipole approximation as can easily be checked by going through the steps leading to Eq. (11) maintaining full retardation and accounting for all electrons.

The pure $3\text{p}_{x}$ and $3\text{p}_{y}$ states of Ar, of course, do not fulfil the requirements for the validity of Eq. (11) and hence the CEP effects observed in this paper are not in general simple rotations. It is, nevertheless, still possible to express momentum distributions for $3\text{p}_{x}$ in terms of momentum distributions for $3\text{p}_{y}$. Symmetry considerations immediately leads to the following formula
\begin{equation}
\frac{\partial^{3} P_{3\text{p}_{y}}}{\partial q_{x} \partial q_{y} \partial q_{z}}(\phi+\pi/2)=\mathbf{R}_{z}(\pi/2)\frac{\partial^{3} P_{3\text{p}_{x}}}{\partial q_{x} \partial q_{y} \partial q_{z}}(\phi),
\end{equation}
where $\phi$ is the CEP. This geometric correspondence is illustrated in Figs. 1 and 2, where we note that $\partial^{3} P_{3\text{p}_{y}}/(\partial q_{x} \partial q_{y} \partial q_{z})(\phi=0)=\mathbf{R}_{z}(\pi/2)\partial^{3} P_{3\text{p}_{x}}/(\partial q_{x} \partial q_{y} \partial q_{z})(\phi=-\pi/2)$.

\section{Conclusion}\label{Conclusion}
We have investigated strong-field ionization of Ar($3\text{p}_{x}$) and Ar($3\text{p}_{y}$) in the presence of a strong circularly polarized laser pulse, by calculating the photoelectron momentum distribution using the TDSE. These systems model the ionization from an oriented target with a single angular nodal plane interacting with a circularly polarized laser pulse, a scenario that could be realized in an oriented molecular target. Our results are compared with results obtained using the LG-SFA and, up to a relatively small rotation of the distribution that is well-understood as a combined field and potential effect \cite{Mig5}, good agreement is observed, emphasizing the LG-SFA as a valuable tool for studying ionization by circularly polarized pulses. Both TDSE and LG-SFA results show distinct signatures of the initial states as well as the temporal shape of the applied pulse. The findings indicate that strong-field ionization by circularly polarized pulses could serve as a probe for revealing nodal structure and also changes that may occur during chemical reactions.

\begin{acknowledgments} This work was supported by the Danish Research Agency (Grant No.
2117-05-0081)
\end{acknowledgments}


\end{document}